\documentclass[reprint, superscriptaddress, amsmath,amssymb, aps, prresearch, longbibliography, floatfix]{revtex4-2}
\usepackage{amsmath}
\usepackage{graphicx}
\usepackage{ulem}
\usepackage{bm}
\usepackage{natbib}
\usepackage{dsfont}
\usepackage{tikz}
\usepackage{epsfig}
\usepackage{feynmf}
\usepackage{blindtext, rotating}
\usepackage{mathtools}
\usepackage{dsfont}
\usepackage{subcaption}
\usepackage{physics}
\usepackage{amsfonts}
\usepackage{xcolor}
\usepackage{ragged2e}
\usepackage{siunitx}
\usepackage{comment}
\usepackage{soul}
\usepackage{empheq}
\usepackage{lipsum}
\usepackage{hyperref} 
\hypersetup{breaklinks=true, colorlinks=true, citecolor=blue, linkcolor=cyan, urlcolor=blue,filecolor=blue}

\usepackage{xcolor} 

\DeclareCaptionJustification{justified}{\justifying}

\DeclareSIUnit{\rad}{rad}

\definecolor{bright_blue}{HTML}{85C1E9}
\definecolor{middle_blue}{HTML}{2E86C1}
\definecolor{dark_blue}{HTML}{1B4F72}

\begin{document}
\title{Apparent violations of the second law in the quantum-classical dynamics of interacting levitated
nanoparticles}

\author{Pedro V. Paraguass\'{u} and Thiago Guerreiro}
 \email{paraguassu@esp.puc-rio.br}
 \email{barbosa@puc-rio.br}
\affiliation{Department of Physics, Pontifical Catholic University of Rio de Janeiro, Rio de Janeiro 22451-900, Brazil}


\begin{abstract}
Random exchanges of energy arise naturally in stochastic systems. As a consequence, apparent violations of the second law of thermodynamics can occur, as it holds true on average. Here we investigate the occurrence of these apparent violations -- termed free lunches -- in a quantum-classical system comprised of levitated nanoparticles exchanging energy via the Coulomb interaction. We consider different initial states for the quantum system, and how these exert work and fluctuations upon the classical particle affecting the probability of free lunches. With that, we initiate the study of hybrid quantum-classical systems through the lens of stochastic thermodynamics. 
\end{abstract}


\maketitle

\section{Introduction}



Across vastly different spatio-temporal scales, physical systems ranging from biological molecules \cite{seifert2018stochastic} to optically levitated nanoparticles \cite{millen2020optomechanics} are subject to stochastic dynamics arising from their interaction with inaccessible environmental degrees of freedom \cite{tome2015stochastic}. To deal with these systems, the framework of stochastic thermodynamics \cite{sekimoto2010stochastic, peliti2021stochastic} defines quantities such as work and heat as random variables \cite{paraguassu2022probabilities, saha2015work, ryabov2013work, chvosta2020statistics, cohen2008properties}. In this context, the usual thermodynamical laws become valid only on average and apparent violations of these laws are possible due to the fluctuating nature of the thermodynamic variables \cite{paraguassu2022probabilities, seifert2005entropy}. For instance, the second law 
\begin{eqnarray}
    \langle w \rangle \geq \Delta F \ ,
\end{eqnarray}
where $ \langle w \rangle $ is the averaged work and $\Delta F$ is the equilibrium free energy, can be violated at the stochastic level. Events which violate the averaged second law are referred to as a \textit{free lunch}, and have recently gained interest both theoretically \cite{paraguassu2022probabilities, salazar2021detailed, merhav2010statistical} and experimentally \cite{barros2024probabilistic, arieltobe}. 

Also lately, levitated nanoparticles have attained prominence as a playground for testing ideas in non-equilibrium physics and stochastic thermodynamics \cite{millen2020optomechanics, debiossac2022non,rademacher2022nonequilibrium,millen2018single,gonzalez2021levitodynamics,raynal2023shortcuts,ciliberto2017experiments,gieseler2015non,gieseler2018levitated}. Due to their high degree of control and isolation, these systems hold great promise for fundamental tests of quantum mechanics \cite{kremer2024all,vanner2013cooling,piotrowski2023simultaneous,gonzalez2021levitodynamics, romero2010toward, tebbenjohanns2021quantum, delic2020cooling,magrini2021real,pikovski2012probing, neumeier2024fast, kamba2023revealing, grochowski2023quantum, weiss2021large} with a number of applications including quantum sensing \cite{moore2021searching,weiss2021large,chaste2012nanomechanical,ranjit2015attonewton,ranjit2016zeptonewton}, tests of collapse models \cite{torovs2017colored,ghirardi1990markov,bassi2013models, millen2020optomechanics}, search of new physics \cite{afek2021limits,afek2022coherent,monteiro2020search,moore2021searching,arvanitaki2013detecting}, probes of the quantum nature of gravity \cite{carlesso2019testing, bose2017spin, van2020quantum, gasbarri2021testing, marshman2020locality,hanif2024testing} and hybrid quantum-classical dynamics \cite{paraguassu2024quantum}, to name just a few. In this work, we set out to connect levitated optomechanics and stochastic thermodynamics via the concept of free lunches, by investigating the violations of the second law in the dynamics of levitated nanoparticles. 

To do so, we will build on \cite{paraguassu2024quantum} and consider the interactions between a quantum and a classical levitated nanoparticle. Our considerations will be quite general and the specific nature of the interaction in itself will not be important, being dependent on the particular implementation; see for instance \cite{rieser2022tunable, rudolph2022force}. In the interest of illustrating the main concepts with a physical example, we will consider the levitated particles interact via the Coulomb force. When the quantum mechanical particle is in a displaced state, such as coherent or squeezed-coherent states, it exerts a deterministic force, therefore performing work on the classical particle through the interaction mechanism. At the same time, quantum fluctuations induce stochasticity in the classical dynamics \cite{caldeira1983path, pelargonio2023generalized, janssen1995nonlinear, feynman2000theory,parikh2021signatures, parikh2021quantum, cho2023graviton, chawla2023quantum}, causing fluctuations susceptible to exhibiting free lunches. Note that the characteristics of the quantum-induced fluctuations are state-dependent: for squeezed states, the noise can be enhanced and exhibit a non-stationary component \cite{parikh2021signatures, paraguassu2024quantum}. The quantum-classical optomechanical dynamics of levitated nanoparticles therefore provides a rich platform for applying stochastic thermodynamics concepts, allowing us to investigate how quantum noise leads to apparent violations of the second law of thermodynamics.





This paper is organized as follows. In Section~\ref{sec2}, we define the quantum-classical optomechanical dynamics used throughout this work. Section~\ref{sec3} defines work and the notion of a free lunch according to stochastic thermodynamics. Section~\ref{sec4} presents general results for the Gaussian work distribution concerning violations of the second law, inherently bound by a 50$\%$ probability of occurrence, while in Section~\ref{sec5} we investigate how these free lunches are affected by classical and quantum noise, by isolating and studying each of these two contributions. In Section~\ref{squeezedcoherent}, we generalize the quantum-classical dynamics and the analysis of free lunches for the case in which the quantum system is in a squeezed-coherent state. We conclude with a discussion of the results and their implications in Section~\ref{discussion}.

\section{Quantum-induced Stochastic optomechanical dynamics}\label{sec2}

Consider two particles trapped by neighboring optical tweezers at a distance $ d $, with displacement coordinates $ x(t) $ and $ y(t) $ and masses $ m $ and $ M $, respectively.  The particles are charged and interact via the Coulomb force. In the limit of small displacements, the interaction potential reads
\cite{rudolph2022force},
\begin{equation}
     V_{e} = -\frac{q_{x}q_{y}}{8\pi \epsilon_{0} d^{3}}(x(t)-y(t))^{2},
\end{equation}
where $ q_{x,y} $ denotes the charges of each particle. The potential $ V_{e} $ contains terms proportional to $ x^{2} $ and $ y^{2} $, whose effect is to shift the natural frequencies of the harmonic traps according to $ \omega_{i} \rightarrow \sqrt{\omega_{i}^{2} - q_{x}q_{y}/4\pi\epsilon_{0}md^{3}} $, and an interaction term $\sim (\hbar g / x_{0}y_{0})\, x(t)y(t) $ with $ g $ being a coupling constant given by,
\begin{eqnarray}
    g = -\frac{q_{x}q_{y}}{4\pi \epsilon_{0} \hbar}\left( \frac{x_{0}y_{0}}{d^{3}} \right).
    \label{langevin}
\end{eqnarray}
where $x_{0} = \sqrt{\hbar/2m\omega_x}$ and $ y_{0} =  \sqrt{\hbar/2M\omega_y}$ are the particles' zero point fluctuations and $ \omega_{x,y} $ are the tweezers' frequencies. We are interested in the dynamics of a classical particle interacting with a quantum mechanical system. Henceforth, we will consider particle $ x $ to be classical, while $ y $ is treated quantum mechanically with a certain initial quantum state. 

The classical-quantum dynamics between levitated nanoparticles has been extensively considered in \cite{paraguassu2024quantum}. In a nutshell, the classical particle follows a stochastic, Langevin-like dynamics given by 
\begin{equation}
    m \ddot x + m\Gamma \dot x + m\omega_{x}^2 x = \zeta(t) + f(t) + \eta(t) , \label{langevin}
\end{equation}
where $ \zeta(t) $ and  $f(t) $ are a quantum-induced noise and deterministic force arising from the interaction with the quantum particle, respectively, and $ \eta(t) $ is a thermal noise describing interactions with additional environmental degrees of freedom, which we will refer to as \textit{classical noise}. This classical thermal noise has zero mean, and correlation given by
\begin{equation}
    \langle \eta(t)\eta(t') \rangle = 2m \Gamma k_BT \delta(t-t'),
\end{equation}
where $T$ is the equilibrium temperature of the environment. Similar to the classical noise, the \textit{quantum-induced noise} $ \zeta(t) $ also has zero mean, with correlation given by
\begin{equation}
    \langle \zeta(t) \zeta(t') \rangle = C \left(\frac{\hbar g}{y_{0}}\right)^2 \cos\left(\omega_y (t-t')\right),
\end{equation}
where $ C $ is an overall factor dependent on the initial state of the quantum particle.

We will be concerned with displaced states of the quantum particle, such as coherent or squeezed-coherent states. For coherent states (that is, in the absence of squeezing) the quantum-induced noise originates from the quantum fluctuations of the ground-state wavefunction, while the deterministic force arises from the amplitude of the displacement, or coherent state amplitude. The deterministic force is,
 \begin{equation}
     f(t) = - 2 \sqrt{n}\left(\frac{\hbar g}{x_{0}}\right)\cos\left(\omega t + \theta\right),\label{detforce}
 \end{equation}
where \(\sqrt{n} = |\alpha|\) is the modulus of the coherent state amplitude, with $ n $ the mean phonon number, and \(\theta\) is the coherent state's phase. 
Later on in Section \ref{squeezedcoherent}, we will also consider squeezed-coherent states. Extension to thermal-coherent states is also possible \cite{paraguassu2024quantum}. 

The dynamics of the classical particle given by Eq.~\eqref{langevin} can be solved for initial conditions $x(0) \equiv x_0$ and $\dot x(0) \equiv  v(0) \equiv v_0$, yielding
\begin{eqnarray}
     x(t) = e^{-\frac{\Gamma}{2}t}\left(x_0 \cos\left(\frac{\Omega t}{2}\right)+\frac{2\left(2v_0+x_0\Gamma\right)}{\Omega}\sin\left(\frac{\Omega t}{2}\right)\right)\nonumber \\ + \frac{2}{m\Omega}\int_0^t dt' e^{-\frac{\Gamma(t-t')}{2}} \sin\left(\frac{\Omega (t-t')}{2}\right)F_{\rm tot}(t'),\nonumber \\
    \label{formal_solution}
\end{eqnarray}
where \(\Omega = \sqrt{4\omega_c^2 - \Gamma^2}\) and \(F_{\rm tot}(t) = \eta(t) + \zeta(t) + f(t)\). From this expression we can calculate all the statistical moments of the position. Moreover, since Eq.~\eqref{langevin} is linear in \(x(t)\) the statistics of the position is Gaussian and all moments are determined by the the mean and variance, which in turn can be calculated from correlation functions over \(F_{\rm tot}(t)\). 


Throughout, we will consider the parameters as shown in Table \ref{parameter_table_coulomb}, adapted from \cite{rieser2022tunable, rudolph2022force, paraguassu2024quantum}.

\begin{table}[t!]
\begin{tabular}{cccc}
\hline \hline
Parameter  &   Symbol    & Units & Value \\ \hline 
 Mass &  $ m $ &   kg    &  $ 10^{-18} $ \\
 Damping rate  &  $ \Gamma $ &   {Hz}    &  $ 10^{-20}$ \\
 Classical Particle frequency  &  $ \omega_{x} $ &  {kHz}    &  $ 2\pi \times 134$ \\
 Quantum Particle frequency  &  $ \omega_y $ &   {kHz}    &  $ 2\pi \times 147$ \\
 Zero point fluctuation  &  $ x_{\rm zpf} $ &   {m}    &  $ 4.1 \times 10^{-12} $ \\
 Coulomb coupling rate &  $ g $ &   {kHz}    &  $ 2\pi \times 51 $ \\

    \hline \hline 
\end{tabular}
\caption{\label{parameter_table_coulomb} System parameters, adapted from \cite{rieser2022tunable, rudolph2022force, paraguassu2024quantum}.}
\end{table}

\section{Elements of Stochastic Thermodynamics}\label{sec3}

Within the framework of stochastic thermodynamics, $f(t)$ can be interpreted as an external deterministic force performing work on the classical particle \cite{bo2019functionals}. We define the work done on the particle in a time interval $t \in [0,\tau]$ \cite{jarzynski1997nonequilibrium},
\begin{equation}
    w[x] = - \int_0^\tau \dot f(t) x(t) dt.
\end{equation}
Note $  w[x] $ defines a functional over \(x(t)\), therefore being itself an unbounded random variable. Its mean is bounded by
\begin{equation}
    \langle w[x]\rangle = W \geq \Delta F\label{secondlaw}
\end{equation}
where \(\Delta F\) is the equilibrium free energy difference over the interval \(\tau\). This bound represents the second law of thermodynamics in its formulation for the work \cite{jarzynski1997nonequilibrium}.

By performing work, we change the internal potential of the system. The second law \eqref{secondlaw} imposes a limitation due to the entropic cost of transitioning from one state to another. For reversible processes we have \(W = \Delta F\), while in general we deal with irreversible processes where \(W > \Delta F\). The entropic cost can be understood by examining the equilibrium free energy,  
\begin{equation}  
    \Delta F = \Delta U - \frac{\Delta S}{\beta},\label{free2}  
\end{equation}  
where \(\Delta U\) is the equilibrium mean energy difference and \(\Delta S\) is the equilibrium entropy difference. For a reversible process $\Delta S = 0$. Therefore, having values less than the free energy would imply $\Delta S < 0$. This does not occur, as the quantity that violates the bound is $w[x]$, not $W$.
For our system, regardless of the form of the deterministic force, the equilibrium free energy difference is given by  
\begin{equation}  
    \Delta F = -\frac{f(\tau)^2}{2m\omega_x^2}.  \label{free3}
\end{equation}  
Note that it depends on the value of the force at the end of the process, the frequency of the classical particle \(\omega_x\), and its mass \(m\). For a derivation of the equilibrium free energy see Appendix \ref{freeapp}.

The idea of apparent violations of the second law is that the work \(w[x]\) can take values where  
\begin{equation}  
    w[x] < \Delta F.  
\end{equation}  
This statistical event is possible because the second law bounds only the mean work. We call this event the \textit{free lunch}, which represents \textit{apparent} violations of the second law. We can calculate the probability of a free lunch $ \mathcal{P}(w < \Delta F)  $, which can be non-zero. 
We are interested in how the state-dependent deterministic force \(f(t)\) in conjunction with quantum-induced noise affects $ \mathcal{P}(w < \Delta F)  $. 

Following the previous discussion on Eq.~\eqref{free2}, observing values where \(W < \Delta F\) would imply \(\Delta S < 0\), which cannot occur. However, at the level of stochastic trajectories we can  define a stochastic entropy \(s = -\ln P(x)\) for which an event where \(w[x] < \Delta F\) corresponds to a decrease in $ s $ \cite{paraguassu2022probabilities}. This stochastic entropy can be negative and is related to the entropy by \(S = \langle s \rangle\) \cite{seifert2005entropy}. 


\section{Gaussian work and free lunch}\label{sec4}

The deterministic force \eqref{detforce} causes oscillations of the classical particle's mean position. 
Due to the linear dependence on position, the work performed by this force is Gaussian distributed.
In view of this, it is useful to highlight general properties of free lunches for a Gaussian distributed work variable. 

The work probability is of the form
\begin{equation}
    P(w) = \frac{1}{\sqrt{2\pi \sigma_W^2}}\exp\left(- \frac{\left(W-w\right)^2}{2\sigma_W^2}\right),
\end{equation}
where $W$ is the mean value given by
\begin{equation}
    W = - \int_0^\tau \dot{f}(t) \langle x(t) \rangle \, dt,
\end{equation}
and $\sigma_W^2$ is the variance,
\begin{equation}
    \sigma_W^2 = \int_0^\tau \int_0^\tau \dot{f}(t) \dot{f}(t') \left[\langle x(t)x(t')\rangle - \langle x(t) \rangle \langle x(t')\rangle\right] \, dt \, dt'.
\end{equation}
For our system, both quantities can be derived analytically with the help of symbolic computation; see appendix \ref{appB} for details on the calculation of the work moments.


The probability of free lunch can be directly computed as,
\begin{eqnarray}
    \mathcal{P}(w < \Delta F) &=& \int_{-\infty}^{\Delta F} P(w) \, dw ,\nonumber \\
     &=& \frac{1}{2}\left(1 + \erf\left(\frac{1}{\sqrt{2}}\frac{\Delta F - W}{\sigma_W}\right)\right).
\end{eqnarray}
Due to the second law \(W \geq \Delta F\), the argument of the error function satisfies 
\begin{equation}
    \chi = \frac{1}{\sqrt{2}} \frac{\Delta F - W}{\sigma_W}\leq 0 \ , 
\end{equation}
introducing an upper bound on the probability of free lunches of \(\mathcal{P} \leq 50\%\) with the maximum value corresponding to \(W = \Delta F\). 
Therefore, for a Gaussian work distribution it suffices to investigate the argument of the error function \(\chi\).

Moreover, we see that we have
\begin{equation}
    \mathcal{P}(w < \Delta F) = \frac{1}{2}\left(1 + \erf\left(-\frac{1}{\sqrt{2}}\frac{W_{\rm irr}}{\sigma_W}\right)\right),\label{freelunch3}
\end{equation}
for $W_{\rm irr} = W - \Delta F$, the mean irreversible work. This accounts for the useful work that one can extract from the system on average \cite{paraguassu2022probabilities}. For a reversible process, it is zero.

We can construct irreversible work as a random variable by defining \( w_{\rm irr} = w - \Delta F \), with its mean given by $W_{\rm irr}= \langle w_{\rm irr}\rangle$, and its variance being the same as the previously defined work variable. That is, \( \sigma_{W_{\rm irr}} = \sigma_W \), since irreversible work is just a constant change in the random variable of the work. This allows us to see the dimensionless ratio 
\begin{eqnarray}
I_{W_{\rm irr}}=W_{\rm irr}/\sigma_W
\end{eqnarray}
as an indicator of the significance of the fluctuations of the irreversible work \cite{jackson2000equilibrium}. The probability of free lunch can be rewritten as
\begin{equation}
    \mathcal{P}(w < \Delta F) = \frac{1}{2}\left(1 + \erf\left(-\frac{1}{\sqrt{2}}I_{W_{\rm irr}}\right)\right),\label{freelunch4}
\end{equation}

\noindent The indicator \( I_{W_{\rm irr}} \) gives the significance of the fluctuations as follows: if \( I_{W_{\rm irr}} \ll 1 \), the fluctuations dominate over the mean, meaning it will be common to observe values that deviate significantly from the mean. Conversely, \( I_{W_{\rm irr}} \gg 1 \) indicates the opposite, where the mean value occurs very often with fewer deviations from the mean.

Regarding the free lunch, if \( I_{W_{\rm irr}} \ll 1 \) we approach the maximum probability of \( 50\% \), since the error function in Eq.~\eqref{freelunch4} will be close to zero. Alternatively, if \( I_{W_{\rm irr}} \gg 1 \), it destroys any chance of a free lunch occurring, because the error function now will give \(-1\). It becomes clear then that more fluctuations allow for a higher probability but always respecting the bound of \(\leq 50\%\).

The above discussion is agnostic to the type of system and noise that we are investigating, only requiring that the work is Gaussian. As a final comment, we highlight that for a non-Gaussian asymmetric distribution the result could be different. Asymmetries can be explored to overcome this Gaussian bound. 


\section{Thermal and quantum noise}\label{sec5}


There are two main sources contributing to work fluctuations: (i) environmental factors, characterized by classical white (thermal) noise and (ii) interaction with the quantum particle, which introduces Gaussian colored noise. Our goal is to explore the characteristics of these noise sources and compare their respective impacts on the occurrence of free lunches. To achieve this, we will examine two specific scenarios. 

First, we will disregard quantum noise. We point out the deterministic force in Eq.~\eqref{detforce} and classical noise can be physically implemented by electrodes and classical control protocols \cite{danger2000efficient,  kremer2024all,kremer2024perturbative} . This scenario represents a purely classical stochastic dynamics. 

We then consider the ideal case where  the classical particle is only affected  by quantum noise.
This approach will enable us to understand how the intrinsic nature of each noise, quantum and classical, affects apparent violations of the second law. Generalization to squeezed quantum states will follow. To this end, we note that the variance can be divided into two parts 
\begin{eqnarray}
\sigma_W^2 = \sigma_\beta^2 + \sigma_\hbar^2 \ ,    
\end{eqnarray}
where $ \sigma_\beta^2 $ denotes the contribution from thermal noise and $\sigma_\hbar^2 $ is due to quantum noise. Note that the mean value of both quantum and classical noises vanish.


\subsection{Irreversible work and reversibility}

As shown in Eq.~\eqref{freelunch3}, violations of the second law are characterized by the mean irreversible work, $W_{\rm irr} = W - \Delta F$. In both the purely classical and ideal quantum scenarios, the irreversible work is identical. What distinguishes both cases are just the variances. Before going into the analysis of free lunches, we first examine the mean irreversible work to better understand the nature of the deterministic force.

\begin{figure}
    \centering
    \includegraphics[width=8.6cm]{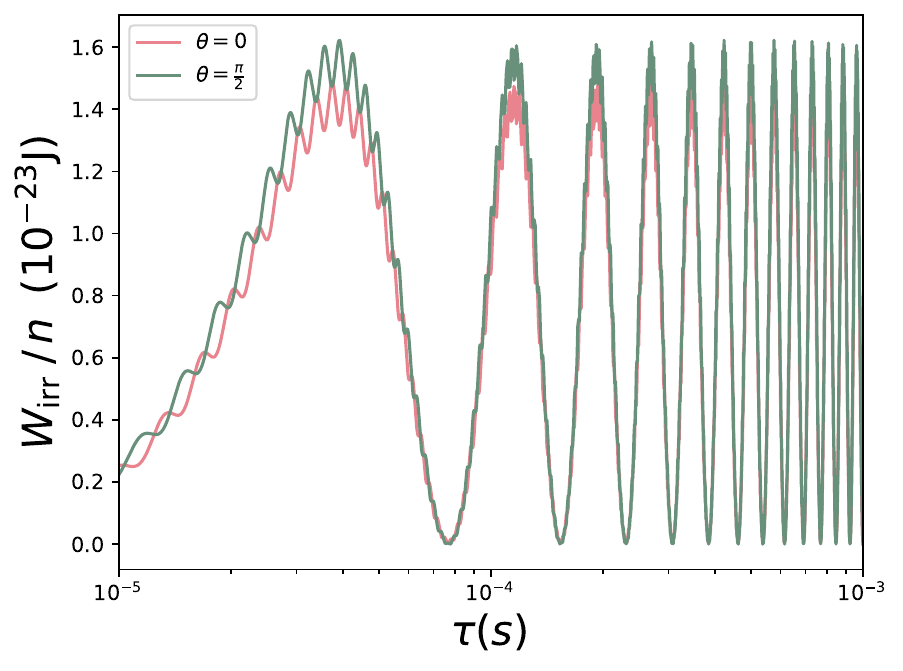}
    \caption{Irreversible work $W_{irr}$ normalized by the number of phonons in the coherent state $n$ as a function of the time interval $\tau$. We consider different values of the coherent phase, namely $\theta = 0 $ and $ \theta = \pi/2$. By varying the time of the process, we can have $W_{irr}=0$, corresponding to a reversible quasi-static process with a higher probability of free lunches.}
    \label{workirr}
\end{figure}

The  irreversible work normalized by the phonon number $n$ is depicted in Figure~\ref{workirr} as a function of the process duration $\tau$. The effects of the coherent state parameters, the phase $\theta$ and the phonon number $n$, are as follows: for the two values of the coherent phase, $\theta = 0$ and $\theta = \pi/2$, we observe a small quantitative difference, with $\theta=0$ yielding larger values than $\theta=\pi/2$. We note this is a small difference, and the phase does not significantly affect the results. In contrast, increasing the phonon number leads to a 
linear increase in the intensity of the work. 

 Note that the mean irreversible work exhibits oscillations with a fixed amplitude throughout the process. An important observation is that $W_{\rm irr} = 0$ for certain time intervals. This occurs due to the oscillatory nature of the deterministic force and depends on the specific combination of parameters listed in Table~\ref{parameter_table_coulomb}. When $W_{\rm irr} = 0$, we have $W = \Delta F$, making the process reversible and consequently quasi-static \cite{jarzynski1997nonequilibrium}. Thus, there exists a combination of time interval and frequency that renders the work quasi-static. A reversible process, $W_{\rm irr} = 0$,  maximizes the probability of 50\% for the occurrence of free lunches. 
 

\subsection{Classical noise}


We consider the classical particle is in contact with a thermal bath at a temperature of 60~K \cite{tebbenjohanns2021quantum}. We assume a deterministic force of the form \eqref{detforce} is exerted at the particle, say by electric feedback,  with $ n = 1, 10, 100 $. This mimics the case in which the quantum particle is in a coherent state with the corresponding mean number of phonons $ n $. 



\begin{figure}
    \centering
    \includegraphics[width=8.6cm]{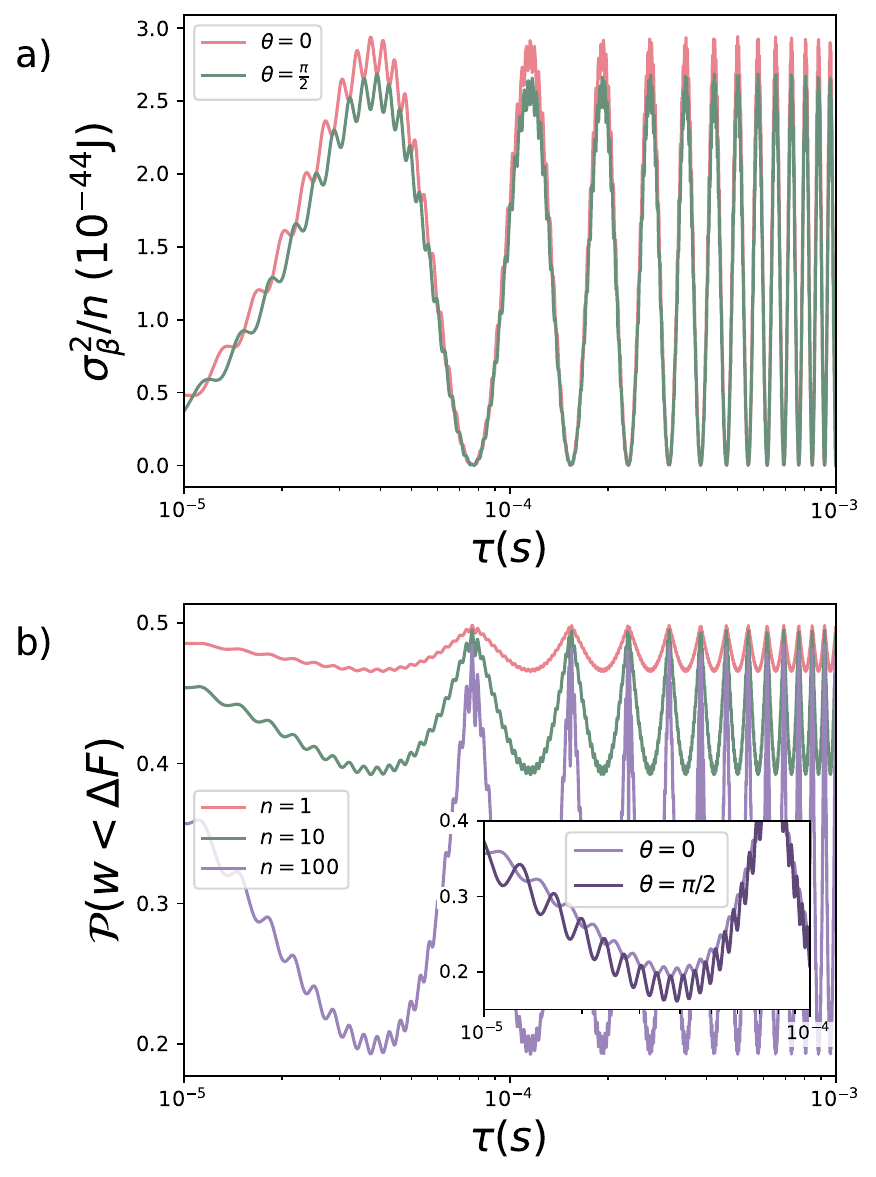}
    \caption{a) Variance for the classical case. We see that when $\sigma_\beta^2=0$ the Gaussian distribution of the work becomes a Dirac delta distribution. b) Probability of free lunch as a function of time $ \tau $, for $n=1,10,100$. Inset: Probability for $n=100$; we see $ \mathcal{P}(W<\Delta F)$ reaches a minimum of $\sim15\%$ for $\theta=\pi/2$, and $\sim 20\%$ for $\theta=0$ and a set of maximums values of 50\%, corresponding to the reversible cases.  }
    \label{varprob}
\end{figure}

A plot of $\sigma_\beta^2$ can be seen in Figure~\ref{varprob}a). We see the effect of the coherent phase is minute, with fluctuations slightly increasing for \( \theta = 0 \) when compared to \( \theta = \pi/2 \). This behavior is mirrored in the probability of the free lunch, as shown in the inset of Figure~\ref{varprob}b). For \( \theta = \pi/2 \), the probability reaches a minimum of approximately \( 15\% \), while for \( \theta = 0 \), it is around \( 20\% \), despite both phases having the same maximum value (due to the quasi-static scenarios). 

Furthermore, extending the time interval does not increase the occurrence of free lunches, since the variance oscillates with a fixed amplitude in Figure~\ref{varprob}a). This result is expected because the significance ratio \( I_{W_{\rm irr}} \) also oscillates with a constant amplitude, which reflects the apparent violations shown in Figure~\ref{varprob}b). Therefore, the maximum and minimum probabilities of free lunches remain fixed in the classical case. As we will see, this behavior  does not occur in the quantum case. We also vary the number of phonons in Figure~\ref{varprob}b), showing that as we increase the phonon number we decrease the free lunch.

Comparing Figures~\ref{varprob}a) and \ref{varprob}b), it is evident that the maximum probability of a free lunch occurs when the variances vanish, which contrasts with the Gaussian behavior discussed in the previous section. The disappearance of the classical variance, shown in Figure~\ref{varprob}a), coincides with the vanishing mean irreversible work displayed in Figure~\ref{workirr}. This raises the question: what happens to the distribution under these circumstances?

We can interpret the simultaneous vanishing of variance and mean of the work by analyzing the irreversible work random variable, which shares the same variance as the usual work. Consequently, the variance of \( w_{\rm irr} \) also approaches zero. Starting from a Gaussian distribution for \( w_{\rm irr} \) with non-zero variance, as the variance goes to zero, the distribution transitions into a delta function. Specifically, \( P(w_{\rm irr}) \rightarrow \delta(w_{\rm irr}) \) in the reversible limit, where \(\langle w_{\rm irr} \rangle = W_{\rm irr} = 0\). This implies that both the variance and the mean vanish under these conditions. For certain time intervals, the distribution effectively deforms into a Dirac delta function.

The reversible case represents the scenario with the maximum probability of free lunches, even when the distributions are no longer Gaussian. Reducing the irreversible work distribution to a delta function, the probability remains \(50\%\), since  
\begin{equation}
    \mathcal{P}(w_{\rm irr} < 0) = \int_{-\infty}^{0} \delta(w_{\rm irr}) \, dw_{\rm irr} = \frac{1}{2} \ ,
\end{equation} 
consistent with our earlier findings. Importantly, this probability corresponds directly to the likelihood of a ``free lunch'', since the condition \( w_{\rm irr} = w - \Delta F < 0 \) describes the same statistical event as \( w < \Delta F \).

\subsection{ Quantum noise}

We now consider the ideal scenario where only quantum noise is present. 
This allows us to analyze in detail how the quantum-classical interaction affects the trajectories of the classical particle, the occurrence of free lunches and the decrease in the system's stochastic entropy.

\begin{figure}
    \centering
    \includegraphics[width=8.6cm]{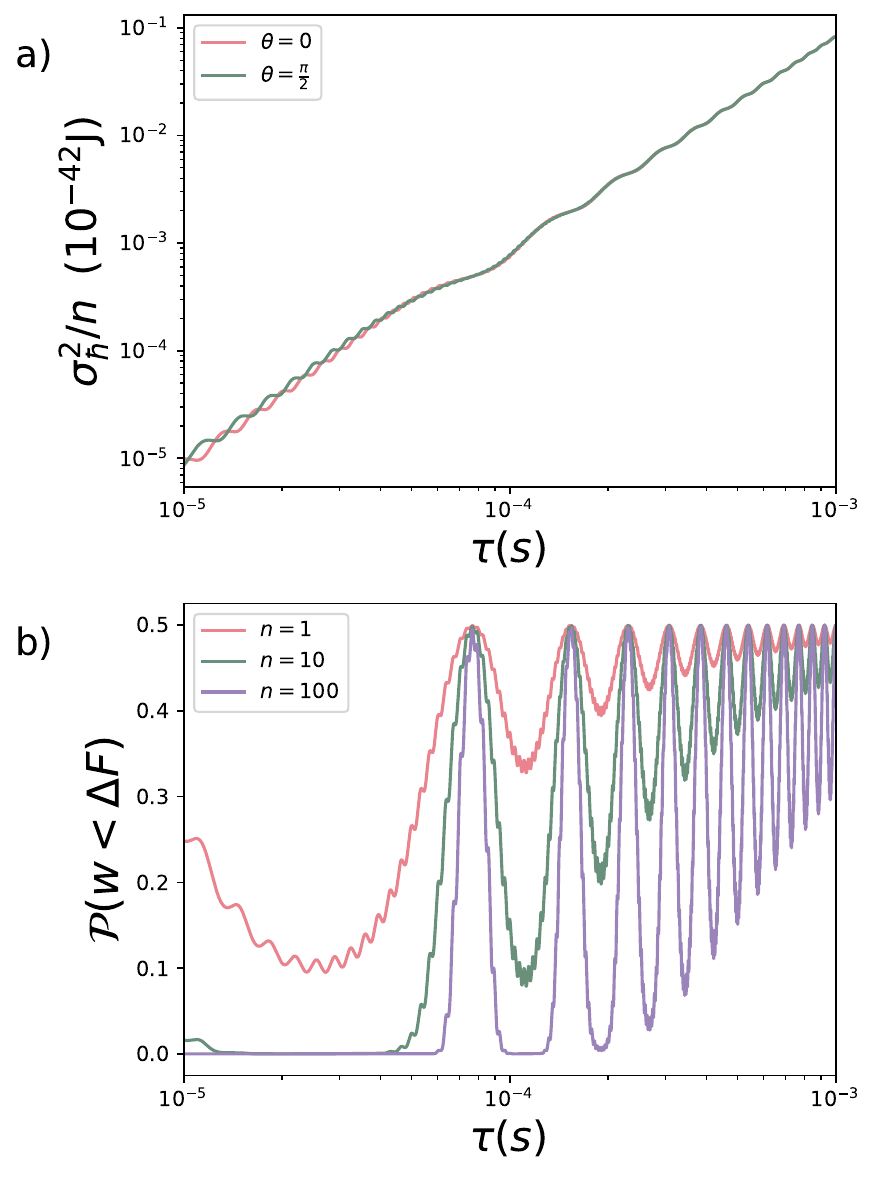}
    \caption{a) Variance for the ideal quantum case divided by the number of phonons $n$. Fluctuations increase as we the time interval of the process gets larger. b) Probability of free lunch as a function of $\tau$ for different values of phonon number. As we increase $n$, the minimum possible values of the free lunch probability decrease. Note that a set of local minima occurs and as the time intervals increase, the probability at these local minima also grows.
}
    \label{qvarprob}
\end{figure}

The variances of the work induced by quantum noise grows with increasing time intervals, as illustrated in Figure~\ref{qvarprob}a). The influence of the coherent phase is also negligible, henceforth we assume \(\theta = 0\). While these fluctuations amplify over time, the irreversible work remains constant. Consequently, the ratio \(I_{W_{\rm irr}}\) decreases, increasing the probability of free lunches. Additionally, the number of phonons amplifies the fluctuations, as \(\sigma_\hbar^2 \sim n\). However, since \(W_{\rm irr} \sim n\), it follows that \(I_{W_{\rm irr}} \sim \sqrt{n}\). Therefore, although the number of phonons increases fluctuations and the significance ratio, it ultimately reduces the probability of free lunches.

In Figure~\ref{qvarprob}b) we observe the probability of free lunches occurring for different phonon numbers and time intervals. Although the maximum probability values remain constant for all phonon numbers, a decrease in the number of phonons results in an increase in the minimum probability, according to \(I_{W_{\rm irr}} \sim \sqrt{n}\). For example, note the region $\Omega_\tau$ =  \(\tau \in [10^{-5}, 10^{-4}]\) s, reducing the phonon number from 100 (purple curve) to 1 (pink curve) causes the probability in that region to rise from \(\min\left[\mathcal{P}_{n=100}(\Omega_\tau)\right] = 0\) to approximately \(\min\left[\mathcal{P}_{n=1}(\Omega_\tau)\right] \sim 10\% \). Moreover, as fluctuations grow with increasing time intervals, these minimum possible values also rise, as illustrated by the shape of the minima in Figure~\ref{qvarprob} b). The probability at the minima increases with longer time durations. For instance, in the region $\Omega_\tau$  for \(n = 1\) (pink curve in Figure~\ref{qvarprob} b)), the minimum probability is approximately 10\%. However, as the interval increases, the minimum probability over the interval increases. Asymptotically, for \(\tau \rightarrow \infty\), the probability tends to oscillate near the upper bound of 50\%, since \(I_{W_{\rm irr}} = W_{\rm irr}/\sigma_\hbar \ll 1\) in this limit.


The occurrence of free lunches is higher in the classical case than in the quantum case, as seen by comparing Figure~\ref{varprob}b) and Figure~\ref{qvarprob}b). For \(n=1\), the classical case oscillates very close to the 50\% bound, while in the quantum scenario, the occurrence of free lunches only oscillates close to the bound for asymptotic time intervals. For \(n=100\), the classical result remains around \( \sim 20\% \) at all local minima, whereas in the quantum case, even though the probability at local minima increases as the time interval grows, for shorter intervals, the minima provide nearly zero probability. Next, we generalize our result to the squeezed-coherent state.


\section{Squeezed-Coherent State}\label{squeezedcoherent}

The quantum particle's coherent state can be squeezed, forming a squeezed-coherent state. Squeezed states are characterized by a reduced standard deviation in a given quadrature operator compared to the ground state standard deviation \cite{bowen2015quantum}. Two quantities defines the squeezed state: the squeezing parameter \(r\) and the phase \(\phi\). For simplicity, we assume \(\phi = 0\). For details on the quantum-classical dynamics induced by squeezed states we refer to \cite{parikh2021signatures, paraguassu2024quantum}.

We now generalize the previous results regarding free lunches to squeezed-coherent states.

\subsection{Dynamics}
The classical stochastic dynamics induced by a quantum particle in a  squeezed-coherent state can be viewed as a generalization of \eqref{langevin}, with the deterministic force and noise correlations modified. 
Specifically, we have the deterministic force
\begin{eqnarray}
    f_{\rm{s}}(t) = - 2 \sqrt{n} \frac{\hbar g}{x_{\rm zpf}} \left( \cosh(r) \cos(\omega t + \theta) \right. \nonumber \\
    \left. + \sinh(r) \sin(\omega t - \theta) \right).
\end{eqnarray}
In addition, squeezing induces a stationary Gaussian noise $ \zeta_{\rm s}(t)$ with correlation
\begin{equation}
    \langle \zeta_{\rm s}(t) \zeta_{\rm s}(t') \rangle = \cosh(2r) \left( \frac{\hbar g}{x_{\rm zpf}} \right)^2 \cos(\omega (t - t')),\label{statnoise}
\end{equation}  
and a non-stationary noise component with 
\begin{equation}
    \langle \zeta_{\rm ns}(t) \zeta_{\rm ns}(t') \rangle = \sinh(2r) \left( \frac{\hbar g}{x_{\rm zpf}} \right)^2 \cos(\omega (t + t')).\label{nonstatnoise}
\end{equation}  
The factor \( \sinh(2r) \) ensures the positivity of the complete noise, which is the sum of the stationary and non-stationary components. Note that for \( r = 0 \), we fall back into the coherent state case. Furthermore, the expression for the free energy \eqref{free3} remains valid, as it is independent of the specific form of the deterministic force.

\subsection{Irreversible work}

\begin{figure}
    \centering
    \includegraphics[width=8.6cm]{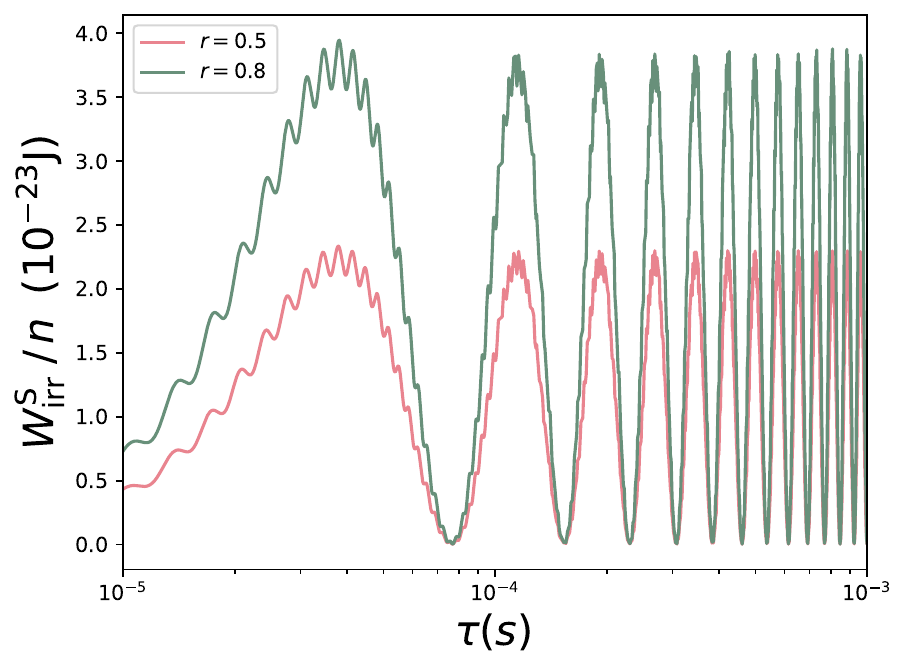}
    \caption{Irreversible work for the squeezed case for two values of the parameter $r$. The amplitude of oscillations increases as we increase the squeezing. As for the coherent state case, the irreversible work becomes null for a set of time intervals, corresponding to reversible processes.}
    \label{worksqueezed}
\end{figure}

In Figure~\ref{worksqueezed} we present the irreversible work for the squeezed-coherent case. Similar to the coherent case, we observe intervals of time where the process becomes reversible, i.e., $W_{\rm irr}=0$, indicating a quasi-static regime.
We compare two values of the squeezing parameter, \( r = 0.5 \) (corresponding to 4.34 dB of squeezing) and \( r = 0.8 \) (corresponding to 6.94 dB). 
We observe that larger squeezing parameters increase the amplitude of the irreversible work, indicating that higher squeezing leads to greater irreversible work.


\subsection{Fluctuations}


We decompose the work fluctuations of the squeezed-coherent case into two contributions: stationary and non-stationary. This separation calls for an analysis of each individual contribution, to understand their respective roles in the occurrence of free lunches.

\begin{figure*}[htbp]
    \centering
    \includegraphics[width=0.95\textwidth]{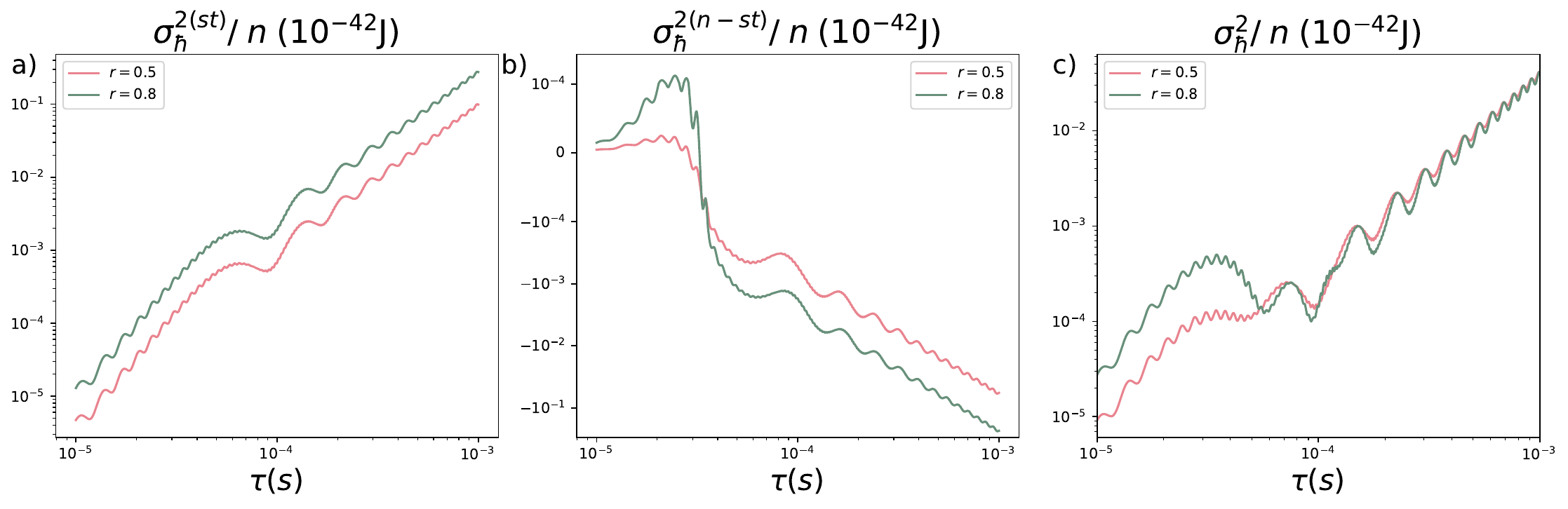}
    \caption{Variance in the squeezed-coherent state. a)Stationary contribution of the variance. b) Non-stationary contribution of the variance. c) Total variance for the ideal quantum squeezing case. A non-monotonic behavior occurs for small values of $\tau$, due to the competition between the stationary and non-stationary contributions.}
    \label{variancessqueezed}
\end{figure*}

In the stationary case, shown in Figure~\ref{variancessqueezed} a), the fluctuations increase as the time interval expands. This effect becomes more pronounced as the squeezing parameter increases. In contrast, the non-stationary contribution, shown in Figure~\ref{variancessqueezed} b) can be negative and decreases as the time interval grows. This negativity is not problematic, as we have $\sigma_\hbar^{2(\text{st})} > \sigma_\hbar^{2(\text{n-st})}$. 
The total variance is shown Figure~\ref{variancessqueezed} c). We see that despite the negative values of the non-stationary part, the total variance always increases as a function of $ \tau $.  



\subsection{Free lunch}

\begin{figure}
    \centering
    \includegraphics[width=8.6cm]{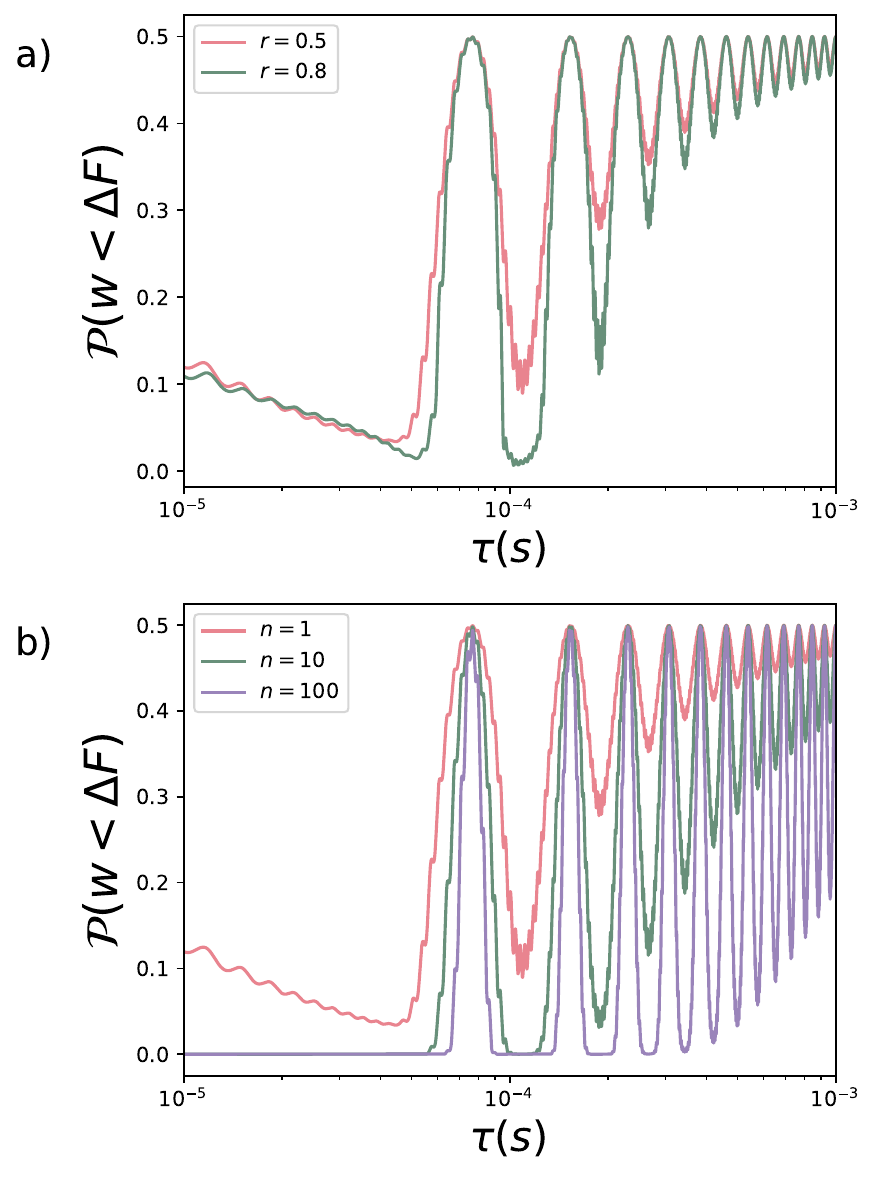}
    \caption{Probability of free lunches in squeezing state. a) Probability as a function of time for different values of the squeezing parameter, with \(n = 1\). Increasing the squeezing parameter raises the probability at the local minima.  b) Probability as a function of time for different phonon numbers, with \(r = 1\). Higher phonon numbers reduce the probability at the local minima. However, the probability at the minima increases as the time interval grows.}
    \label{probsqueezed}
\end{figure}

As in the previous section, the free lunch probability remains Gaussian, as squeezing does not alter the statistical nature of the work. In Figures~\ref{probsqueezed}a) and b) we present the probability of free lunches as a function of squeezing parameter and number of phonons, respectively. Similarly to the quantum case described in Section \ref{sec5}, the increase in fluctuations with the duration of the process leads to a higher probability of free lunches. This probability stabilizes asymptotically and shows minor oscillations near the upper bound of 50\%.

As shown in Figure~\ref{probsqueezed} a), the probability in the local minimum for the squeezed-coherent state is lower than that observed in the quantum case (Figure~\ref{qvarprob} b)). While squeezing amplifies the fluctuations -- despite the reduction due to the non-stationary contribution -- the irreversible work also increases. Consequently, the significance ratio \( I_{W_{\rm irr}} \) is bigger for the squeezed coherent state than for the quantum case. Specifically, as the squeezing parameter increases from 0.5 to 0.8, the probability of free lunches occurring at the local minima diminishes.

At first, this result may seem counterintuitive, as squeezing amplifies fluctuations. However, the irreversible work is proportional to \(\cosh^2(r)\), while the variance is proportional to \(\cosh^{3/2}(r)\), resulting in the significance ratio being proportional to \(\cosh^{1/2}(r)\). Since \(\cosh^{1/2}(r)\) increases with \(r\), the occurrence of free lunches decreases as \(r\) increases.

The effect of the number of phonons is illustrated in Figure~\ref{probsqueezed} b), where the squeezing parameter is fixed at \( r = 0.5 \). As shown, increasing the number of phonons eliminates any possibility of a free lunch occurring during the initial time intervals. In particular, for \( n = 100 \) (purple curve in Figure~\ref{probsqueezed}), within the region  \(\Omega_{\tau}= \tau \in [10^{-5}, 5 \times 10^{-5}] \, \mathrm{s} \), the probability \( \mathcal{P}(w < \Delta F) \) is zero. However, as the process duration increases, the probability of free lunches re-emerges. This is due to the minimum probabilities in the increasing ranges continuing to increase over time, consistent with the results discussed earlier.

\section{conclusion}\label{discussion}

In this work, we utilized concepts from stochastic thermodynamics, such as the notion of work as a random variable and the occurrence of apparent violations of the second law, free lunches, to investigate how noise arising from the interaction between a classical and a quantum system affects the occurrence of trajectories where the stochastic entropy decreases. We also investigated how such apparent violations depend on the state of the quantum system, notably for coherent and squeezed-coherent states. 
Overall, find that a thermal classical noise exhibits more free lunches than quantum-induced noise. While squeezing increases quantum fluctuations, the occurrence of free lunches is lower in the squeezed-coherent state compared to coherent states.

The work done on the classical particle due to the quantum particle is a Gaussian random variable, which imposes an upper limit of 50\% on the probability of free lunches. 
This bound is characteristic of the behavior of a Gaussian random variable, which presents a symmetric distribution with zero skewness. The only way to achieve a probability $\mathcal{P} > 50\%$ with $W \geq \Delta F$ is through an asymmetric distribution. An example was reported in \cite{barros2024probabilistic}. It is worth noting that our study was limited to Gaussian states; therefore, considering different dynamics could affect the conclusions presented here. Future investigations and generalizations in this direction are the subject of further work.

Apparent violations can be exploited in thermal machines, where increasing the probability of free lunches consequently raises the probabilities of achieving higher efficiencies. Moreover, the increase in free lunches could assist in experimentally verifying the Jarzynski equality, as a sufficient number of trajectories where $ w < \Delta F $ are required. Additionally, the emergence of colored noise due to quantum-classical interaction opens up new theoretical avenues for investigating the dynamics and thermodynamics of such systems.

\appendix

\section{Equilibrium Free Energy}\label{freeapp}

To calculate the equilibrium free energy, $\Delta F = \Delta U- \beta^{-1}\Delta S$, we need the potential $U$ and the entropy $S$ in equilibrium. First we notice that the potential energy difference in equilibrium will be given by $\Delta U = \langle \Delta u \rangle_0$, where $\langle \dots\rangle_0$ is the equilibrium average, and $\Delta u = u_f-u_0$ is the random variable potential energy.

At $t=0$, we consider that our system is in equilibrium, not interacting with the quantum particle, therefore the initial random variable potential energy is $u_0=\frac{1}{2}m\omega_c^2 x_0^2$. After interacting, the quantum particle exerts a deterministic force, and in the end of the process, the final random variable potential energy will be 
$u_f = \frac{1}{2}m\omega_c^2 x_\tau^2 -f(\tau)x_\tau $. The average $\langle \dots\rangle_0$ is obtained by considering equilibrium of the variable $x_0$ and $x_{\tau}$. The equilibrium distribution of both quantities are
\begin{equation}
    P_0(x_0) \sim \exp\left(-\beta \frac{m \omega_c^2}{2}x_0^2\right),
\end{equation}
for the initial position, and
\begin{equation}
    P_\tau(x_\tau) \sim \exp \left(-\beta \left(\frac{1}{2}m\omega_c^2 x_\tau^2 - f(\tau) x_\tau\right)\right).
\end{equation}

Meanwhile, $\Delta S$ is the difference in Shannon entropy between two equilibrium states, where $S=-\int p(x)\log(p(x))dx$. It can be understood as the mean of the stochastic entropy, $s=-\log p(x)$, where in equilibrium $S=\langle s \rangle_0$ Therefore, we can calculate it using the above distributions.  

To compute the equilibrium free energy, we take into account the equilibrium position distributions, calculating $\langle \Delta u\rangle_0$, and $\langle \Delta s\rangle_0$, and putting all together in $\Delta F = \Delta U- \beta^{-1}\Delta S$ give us
\begin{equation}
    \Delta F = - \frac{f(\tau)^2}{2 m \omega_c^2}.\label{freenergy}
\end{equation}

\section{Work moments}\label{appB}

To calculate the work moments, the mean and the variance, we need the first moments of the position. We take the average over Eq.~\eqref{formal_solution}, giving
\begin{eqnarray}
    \langle x(t)\rangle =\nonumber\\  e^{-\frac{\Gamma}{2}t}\left(\langle x_0 \rangle \cos\left(\frac{\Omega t}{2}\right)+\frac{2\left(2\langle v_0\rangle+\langle x_0\rangle\Gamma\right)}{\Omega}\sin\left(\frac{\Omega t}{2}\right)\right)\nonumber \\ + \frac{2}{m\Omega}\int_0^t dt' e^{-\frac{\Gamma(t-t')}{2}} \sin\left(\frac{\Omega (t-t')}{2}\right)f(t'),\nonumber \\
\end{eqnarray}
where initially the system is in equilibrium, therefore $\langle x_0 \rangle$ and $\langle v_0 \rangle$ are calculated with a equilibrium distribution.

Following the same idea, we multiply $x(t)$ by $x(t')$ and take the average over the product. This will give the correlation $\langle x(t)x(t')\rangle$, allowing us to calculate $\langle x(t)x(t')\rangle - \langle x(t)\rangle \langle x(t')\rangle$ straightforwardly.

With both, $\langle x(t) \rangle$ and $\langle x(t)x(t')\rangle - \langle x(t)\rangle \langle x(t')\rangle$ in hands, the mean and the variance of the work will be given respectively by direct integration of 
\begin{equation}
    W = - \int_0^\tau \dot{f}(t) \langle x(t) \rangle \, dt,
\end{equation}
\begin{equation}
     \sigma_W^2 = \int_0^\tau \int_0^\tau \dot{f}(t) \dot{f}(t') \left[\langle x(t)x(t')\rangle - \langle x(t) \rangle \langle x(t')\rangle\right] \, dt \, dt'.
\end{equation}
These expressions are long and complicated, but can be calculated directly by integration. Any symbolic computation software can handle the calculations easily.

\section*{ACKNOWLEDGEMENTS}
P.V.P acknowledges M. I. Martins and S. Aires for their useful help in the preparation of the manuscript, and the Funda\c{c}\~ao de Amparo \`a Pesquisa do Estado do Rio de Janeiro (FAPERJ Process SEI-260003/000174/2024). T.G. acknowledges the Coordena\c{c}\~ao de Aperfei\c{c}oamento de Pessoal de N\'ivel Superior - Brasil (CAPES) - Finance Code 001, Conselho Nacional de Desenvolvimento Cient\'ifico e Tecnol\'ogico (CNPq), Funda\c{c}\~ao de Amparo \`a Pesquisa do Estado do Rio de Janeiro (FAPERJ Scholarship No. E-26/200.252/2023 and E-26/202.762/2024)  and Funda\c{c}\~ao de Amparo \`a Pesquisa do Estado de São Paulo (FAPESP process No. 2021/06736-5). This work was supported by the Serrapilheira Institute (grant No. Serra – 2211-42299 ) and StoneLab.

\bibliography{name}

\end{document}